\documentstyle[12pt,aasms4]{article}

\righthead{Nonthermal EUV/X-Ray Emission of Coma}
\lefthead{Atoyan \& V\"olk}

\begin{document}

\title{Implications of a Nonthermal Origin of the Excess EUV \\
Emission from the Coma Cluster of Galaxies}

\author{A. M. Atoyan}
\affil{Max--Planck--Institut f\"ur Kernphysik, P.O. Box 103980,
D-69029 Heidelberg, Germany}
\affil{Yerevan Physics Institute, Alikhanian Broth. 2, 375036 Yerevan,
Armenia}
\authoremail{Armen.Atoyan@mpi-hd.mpg.de}

\and

\author{H. J. V\"olk}
\affil{Max--Planck--Institut f\"ur Kernphysik, P.O. Box 103980,
D-69029 Heidelberg, Germany}
\authoremail{Heinrich.Voelk@mpi-hd.mpg.de}

\begin{abstract}
The inverse Compton (IC) interpretation  of the 
excess EUV emission, that was recently  reported from several 
clusters of galaxies,  
suggests that the amount of relativistic electrons in the intracluster
medium is highly significant, $W_{\rm e}> 10^{61}\,\rm erg$. 
Considering Coma as the prototype galaxy 
cluster of nonthermal
radiation, with synchrotron and IC fluxes measured in the radio and EUV
regions, and possibly also in the hard X-ray region, we discuss implications 
of the inverse
Compton origin of the EUV fluxes in the case of low intracluster 
magnetic fields of
order $0.1\,\rm \mu G$ as required for the IC interpretation of the 
observed excess hard X-ray flux,  and in the case of high fields of
order $1\,\rm \mu G$ as suggested by Faraday rotation measurements. 
Although for such high intracluster fields the excess hard X-ray fluxes 
will require
an explanation other than by the IC effect, we show that the excess   
EUV flux can be explained by the IC emission of a  
`relic' population of electrons driven into the incipient 
intracluster medium at the
epoch of starburst activity by galactic winds, 
and later on reenergized by adiabatic compression and/or large-scale
shocks transmitted through the cluster as the consequence of more
recent merger events.    
Radiative cooling will naturally produce a sharp
cutoff in the spectrum  of these relic electron population, which
is required, in the case of $\mu$-Gauss fields, in order to avoid a 
contradiction with the observed 
radio fluxes. For high magnetic fields $B \geq 1\,\rm \mu G$
the interpretation of the radio fluxes of Coma requires a second population 
of electrons injected recently. They can be explained as  
 secondaries produced by a population of 
relativistic protons.
We calculate the fluxes of $\gamma$-rays to be expected
in both the low and high magnetic field scenarios, and discuss possibilities
to distinguish between these two principal options by future $\gamma$-ray
observations.
\end{abstract}


\section{Introduction}

One of the significant characteristics of galaxy clusters 
is their nonthermal particle content. 
A clear signature of the presence of  
relativistic electrons, and by implication also of relativistic protons,
in the intracluster medium (hereafter ICM) is the detection of
a diffuse cluster-scale radio emission from a number of these objects
(\cite[and references therein]{Gio99}). The radio
fluxes give an important information on the energy spectra of the
electrons in the ICM, but being dependent on the ICM magnetic fields, they
do not yet allow unambiguous estimates of the overall energy content
of relativistic electrons. 
More accurate estimates can be derived  
if one interprets the `excess' diffuse EUV emission  
reported from some clusters 
(\cite{Lie96,Mit97,Bow98}; but see also \cite{arabad})
in terms of inverse Compton (IC) 
upscattering of 2.7\,K microwave background radiation by 
electrons with energies of a few 100\,MeV (\cite{Hwa97,Sar98,Ens98}). 
This typically requires an  
overall energy in these electrons 
of about $5\times 10^{61}\,\rm erg$. 
Depending on the spectra of relativistic electrons
at lower energies, and on the ratio of the relativistic protons to electrons,
the total energy in cosmic rays (CRs) might be then as high
as $W_{\rm CR }\sim (10^{62}-10^{63})\,\rm erg$. 

Thus, CRs may contain a significant fraction of the overall kinetic 
energy in the 
ICM and could be dynamically important. Because the energy loss 
time of the dominant hadronic CR particles in the ICM typically exceeds the 
Hubble time, and since their confinement time generally exceeds the cluster
lifetime (\cite{Vol96,Ber97,Col98}), the nonthermal component of 
galaxy clusters 
should contain valuable information about the evolution of these
largest gravitationally bound systems of the Universe. 
  
In this paper we study in detail the broad band nonthermal emission
of the Coma cluster, considered as a prototype,
with nonthermal radiation fluxes reported  
both in the radio and the EUV regions. In the Coma cluster 
the `excess' (over a single-temperature plasma emission) flux  
is firmly established also in the region of hard X-rays by  
\cite{Fus99} (1999) and \cite{Rep99} (1999). The 
interpretation by these authors of this hard X-ray emission in terms of IC 
radiation of electrons which are also 
responsible for the radio fluxes at low frequencies, 
suggests that the ICM magnetic field in Coma is about
$0.1-0.2 \,\rm \mu G$. In section 2 below we 
summarize the results following from the hypothesis that, along with 
the excess EUV flux, the excess hard X-ray flux also has an IC 
origin,
and calculate the fluxes of IC radiation to be expected in the 
$\gamma$-ray region. 

However, it is quite possible that the ICM magnetic field
in Coma is at a level of $\geq 1\,\rm \mu G$, as deduced from
Faraday rotation and depolarization measurements on background radio sources
(\cite{Kim90,Fer95}).
Although an IC origin of the excess hard X-ray flux in that case is excluded,
we show  in section 3 that 
an IC origin of the  excess EUV flux is still possible.
We demonstrate that a rather unusual energy distribution of the relativistic
electrons, with a sharp cutoff above a few 100 MeV (that is needed to 
comply with the data both in EUV and the radio region), 
can be formed during the cosmological evolution of the Coma cluster.
In particular, these electrons could be the relics of the early starburst 
activity in the Coma galaxies which fed the incipient ICM with 
relativistic particles and 
magnetic fields via their galactic winds (\cite{Vol99}). 
In section 4 we calculate the $\gamma$-ray fluxes to be expected in the 
 high magnetic field case and discuss whether it would be possible
 to distinguish between high and low magnetic field scenarios by future
 $\gamma$-ray observations.

\section{IC interpretation of the excess hard X-rays}

Radio fluxes from the Coma cluster are detected from 10.3\,MHz to 2.7\,GHz
(\cite{Bri68,Hen89,Kim90,Gio93}),
and at 4.85\,GHz an upper flux limit was reported by Kim et al. (1990).
 The differential energy fluxes 
$J_{\nu} \propto \nu^{-\alpha_{\rm r}}$ 
between
30.9 MHz and 1.4 GHz are well fitted with a power-law index 
$\alpha_{\rm r} =1.16$.  The data at 2.7 and 4.85 GHz fall
below this extrapolation, so that the `best fit' single power-law index
would then correspond to $\alpha_{\rm r} =1.36$ (\cite{Sch87}).
Note however that the reported sharp decline of the radio fluxes above 1.4 GHz 
can also be explained as an instrumental effect (\cite{Deiss97}). 

The diffuse excess (over the thermal) EUV radiation between 70\,eV and
$\simeq 200 \,\rm eV$ has a mean power-law index $\alpha_{\rm euv}\sim 0.75$
(Lieu et al. 1999). The power-law index of the diffuse excess X-rays
is rather uncertain and lies most probably within the range of 
$\alpha_{\rm x}\simeq (1-2)$ (see Fusco-Femiano et al. 1999; Rephaeli et
al. 1999), which includes the values of the 
spectral indices $\alpha_{\rm r}$ of the measured radio fluxes.
The latter circumstance becames important for an IC  
interpretation of the excess X-ray flux of Coma because this radiation
should be then produced by the same
population of electrons which are responsible for the synchrotron emission
at radio frequencies.   

The characteristic frequency of synchrotron radiation of electrons with
Lorentz factor $\gamma$ is 
$\nu \simeq B_{\mu \rm G} \, \gamma^2\,\rm Hz$ 
where $ B_{\mu \rm G}\equiv B/1\,\mu {\rm G}$ (\cite{Gin79}).  
Thus the production of synchrotron radiation  
in the region from 10 MHz to 1.4 GHz  requires that the
energy distribution of the electrons $N(\gamma)\propto
\gamma^{-\alpha_{\rm e}}$  extends with an index $\alpha_{\rm
e}=1+2\alpha_{\rm r}\approx (3.2-3.6)$ from $\gamma_{\rm r,1} 
\simeq 3\times 10^3\, B_{\mu \rm G}^{-0.5}$ to
$\gamma_{\rm r,2} \simeq 4 \times 10^4\, B_{\mu \rm G}^{-0.5}$. 

For the IC process the mean energy of photons produced in the 
 2.7\,K background radiation field with the mean photon energy
$\epsilon_0\simeq 3\,k T$ is $\epsilon \simeq (4/3)
\epsilon_0 \,\gamma^2 \simeq 8.7\times
10^{-4}\,\gamma^2 \,\rm eV$. For the production of the X-ray photons from 
$20\,\rm keV$ to $80\,\rm keV$ one therefore needs electrons with energies
between $\gamma_{\rm  x, 1}\simeq 5\times 10^3$ and 
$\gamma_{\rm x, 2}\simeq 10^4$. Because the latter actually 
corresponds to the lower edge $\gamma_{\rm r,1}$ of the electrons responsible
for the observed radio emission  
(in magnetic fields of order $0.1\, \mu \rm G$), the spectral index
of the X-ray emitting electrons should be close to $\alpha_{\rm e}$
as derived from radio data. 
 
The IC origin of Coma's EUV spectrum from 70\,eV to 
200\,eV implies a spectral index $\alpha_{\rm e}\sim
2.5$ for the electrons in a rather narrow energy region 
around $\gamma_{\rm euv} \simeq 300$ (Sarazin \& Lieu 1998). A gap by a 
factor of about $10$ between 
 $\gamma_{\rm euv}$ and  $\gamma_{\rm r,1}$ allows then a scenario 
with a single population of power-law electrons where radiative losses
induce a break in $\alpha_{\rm e}$ 
somewhere in that gap. Thus assuming
a source function of electrons $Q(\gamma)\propto \gamma^{-\alpha_{\rm inj}}$
 in the ICM with 
$\alpha_{\rm inj}=2.3$, the IC emission of those electrons at low energies
will give $\alpha_{\rm euv}=0.65$, which is acceptably 
close to the value 0.75. After the break, the electron distribution steepens 
to $\alpha_{\rm e}=1+\alpha_{\rm inj}=3.3$ required for interpretation of 
the radio fluxes. 

Such a break in the energy spectrum  can be  produced by radiative
losses of electrons if they are  injected 
into the ICM on time scales $\Delta t_{\rm inj} \geq 10^9\,\rm yr$. For the 
radiative energy
loss rate due to IC (in the 
Thompson limit) and synchrotron processes (e.g. Ginzburg 1979)
\begin{equation}
P_{\rm rad}(\gamma)=-\frac{{\rm d}\gamma}{{\rm d}t} 
=\frac{4\, \sigma_{\rm T}}{3\, m_{\rm e} c}
\left( w_{\rm mbr} + \frac{B^2}{8\pi}\right) \gamma^2\; ,
\end{equation}
the characteristic energy loss time is  
\begin{equation}
t_{\rm rad}\approx 2.4\times 10^{12} [(1+z)^4 +0.1 B_{\mu \rm G}^{2}]^{-1} 
\,\gamma^{-1} \;  \rm yr \; , 
\end{equation} 
where we take into account the cosmological evolution of the 
energy density of the cosmic microwave background radiation (MBR),  
$w_{\rm mbr}\propto (1+z)^4$ at the epoch $z$.
For magnetic fields $B< 3\,\rm \mu G$ the radiative losses 
are dominated by the IC process. Using  Eq.\,(2), from the condition 
$t_{\rm rad}(\gamma)\simeq \Delta t_{\rm inj}$ we find that 
the radiative break would appear in the `proper' place 
($\gamma_{\rm br} \sim 10^3 $) if relativistic electrons 
were injected into the ICM recently, during times 
$t_{\rm inj}\sim (1-3)\times 10^9\,\rm yr$ (this corresponds to 
$z_{\rm inj}\leq 0.2$). Note that the 
relativistic electrons need to be produced/accelerated in the ICM 
{\it continuously} during that time period, because in the case of an 
`impulsive' injection the radiative losses induce a cutoff, 
and not a break,  above $\gamma_{\rm br}$. 

\begin{figure}
\plotone{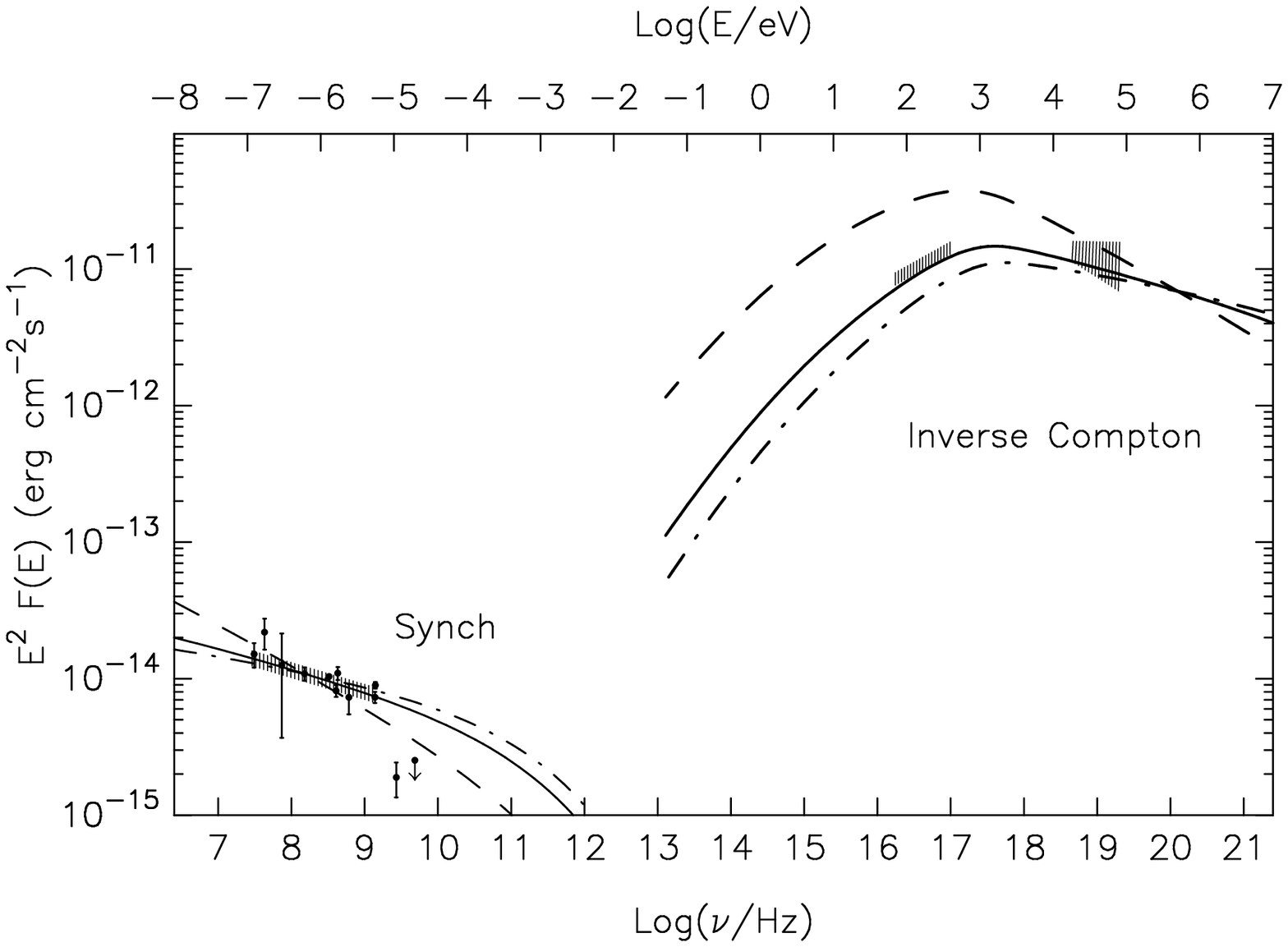}
\figcaption{The synchrotron and IC fluxes calculated for 
the magnetic field in Coma cluster $B=0.12 \,\rm \mu G$, assuming 
stationary injection of relativistic electrons during the last  
$\Delta t_{\rm inj} = 3\times 10^9\,\rm yr$, and 3 different power-law
indices for the source function $Q(\gamma)$ in Eq.(3): 
$\alpha_{\rm inj}=2.2$ (dot-dashed curves), $\alpha_{\rm inj}=2.3$  
(solid), and $\alpha_{\rm inj}=2.6$ (dashed). An exponential cutoff
energy $\gamma_{\rm c}=2\times 10^6$ is assumed. The source function 
is chosen so that the radio spectrum is normalized to the observed 
flux at 400\,MHz. The compilation of radio data and the slope of
the hatched region from 30\,MHz up to 1.4\,GHz, corresponding to a power-law
index $\alpha_{\rm r}= 1.16$ for the differential energy flux 
$J(E)=E F(E)$, are taken from Deiss et al. (1997). The hatched regions
in the EUV and X-ray domains correspond to the fluxes observed by
Lieu et al. (1999), and Rephaeli et al. (1999) and 
Fusco-Femiano et al. (1999), respectively. \label{fig1}}
\end{figure}

\placefigure{fig1}

In Fig.~1 we show the results of calculations when  
the same $B = 0.12 \,\rm \mu G$ and $\Delta t_{\rm inj}= 3 \,\rm Gyr$ 
are assumed for 3 different power-law indices $\alpha_{\rm inj}$ of the electron 
injection spectrum
\begin{equation}
Q(\gamma)\propto \gamma^{-\alpha_{\rm inj}} \exp (-\gamma/\gamma_{\rm c}),
\end{equation}
where $\gamma_{\rm c}$ defines the characteristic maximum energy
of the accelerated particles. The synchrotron fluxes are
normalized to the radio flux from the  Coma cluster observed 
at $400\,\rm MHz$. For the assumed magnetic field, 
the IC `counterpart' of this normalization point in Fig.~1 is found at
the photon energy $\epsilon \approx 1\,\rm MeV$, and therefore for different
indices $\alpha_{\rm inj}$ the IC fluxes predicted in the X-ray region
are somewhat different. Because the synchrotron flux at a given frequency
depends on the overall number of radio electrons and magnetic field as   
$L_{\rm syn}\propto N_{\rm e} B^{(1+ \alpha_{\rm e})/2}$,
the X-ray fluxes can be increased or decreased
assuming slightly lower or higher magnetic fields. Thus for the injection spectrum
with $\alpha_{\rm inj}=2.2$ a field somewhat smaller than the 
$B\approx 0.12\,\rm \mu G$ assumed in Fig.~1 would be needed to explain
the observed fluxes both in the EUV and 
in the X-ray regions. In the case of a steep spectrum with 
$\alpha_{\rm inj}=2.6$ an increase of the field to $B\approx 0.15\,\rm \mu G$
could reduce the IC flux to within the range of the observed `excess' fluxes 
in the X-ray region. But, in addition, we have to assume a more recent time
for the injection of relativistic electrons in the ICM -- in order to induce 
the radiative break at higher energies and to explain also the excess EUV flux.

\begin{figure}
\plotone{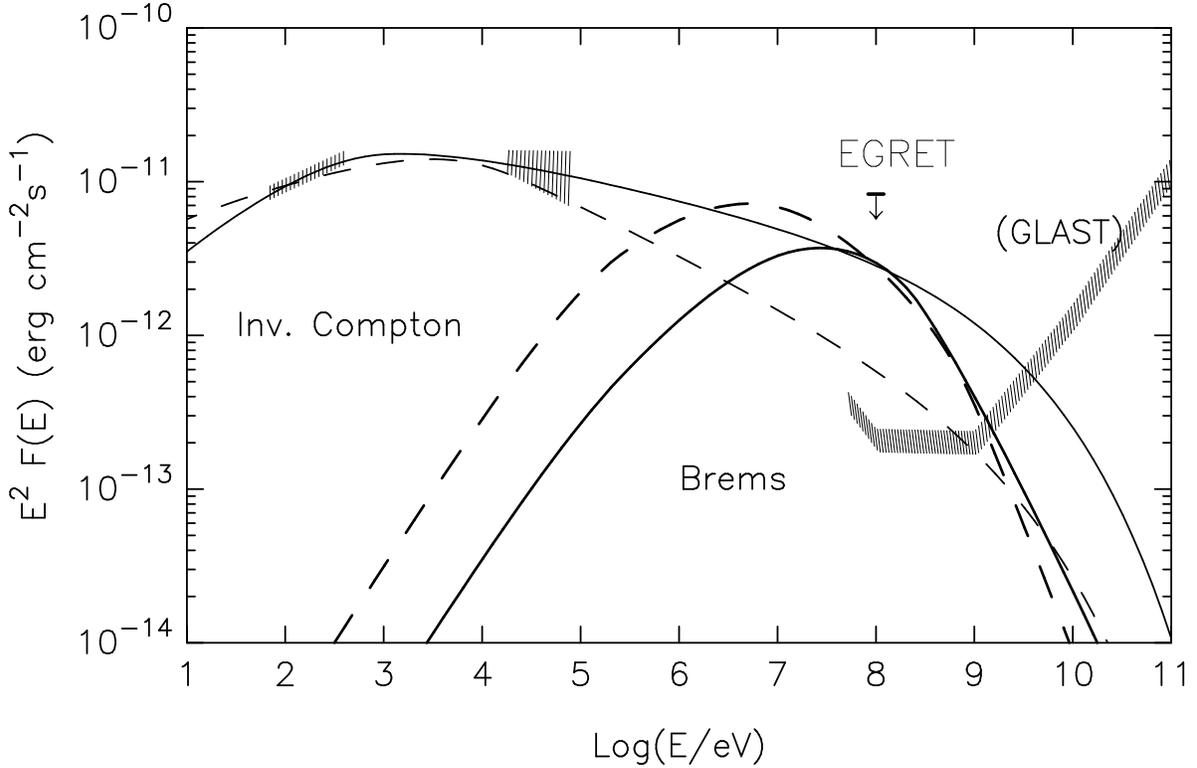}
\caption{ The bremsstrahlung and IC radiation fluxes calculated 
in the case of injection of relativistic electrons with 
$ \alpha_{\rm inj} =2.3$ during the last $\Delta t_{\rm inj}=3\,\rm Gyr$ assuming 
$B=0.1 \,\rm \mu G$ (solid curves), and $ \alpha_{\rm inj}=2.6$, 
$\Delta t_{\rm inj}=1\,\rm Gyr$ assuming $B=0.15 \,\rm \mu G$ (dashed curves).
A mean gas density $n_{\rm gas}=10^{-3}\,\rm cm^{-3}$ in the ICM is assumed.    
In the $\gamma$-ray region, the expected flux sensitivity of the 
GLAST detector (from Bloom 1996) and the upper flux limit of EGRET
(Sreekumar et al. 1996) are also shown. \label{fig2}}   
\end{figure}

\placefigure{fig2}

Thus, uncertainties in the reported fluxes and spectral indices of the radiation
in different energy bands allow some latitude in the predictions 
of the magnetic field in the Coma cluster. In Fig.~2 we show the spectra
of IC radiation calculated for two different sets of the parameters
$B$, $\alpha_{\rm inj}$, and $\Delta t_{\rm inj}$. Note that the value 
of $B\approx 0.15\,\rm \mu G$ needed in the case of a steep radio spectrum
with $\alpha_{\rm r}=1.3$ (i.e.  $\alpha_{\rm inj}=2.6$) agrees well 
with the magnetic field deduced by Fusco-Femiano et al. (1999) from comparison
of the radio fluxes with the excess X-ray flux detected by the BeppoSAX
detector. A further increase of the magnetic field, up to $B\leq 0.2\, \rm
\mu G$ (Rephaeli et al. 1999), would be possible only for a smaller flux of  
the IC X-rays.
 
The total energy in relativistic electrons which is needed for the interpretation
of the excess EUV and X-ray fluxes in Fig.~2 is equal 
to $W_{\rm e}=6.3\times
10^{61}\rm erg$ and  $W_{\rm e}=1.8 \times 10^{62}\rm erg$ for 
$\alpha_{\rm inj}=2.3$ and $\alpha_{\rm inj}=2.6$, respectively. 
For the angular size of the EUV excess
${\rm FWHM} \approx 19^\prime \times 13^\prime$ (Bowyer \& Bergh\"ofer 1998)
the mean electron energy density is of order $ 1\,\rm eV/cm^3$, and
exceeds the magnetic energy density 
$w_{\rm B}=2.5\times 10^{-4}(B/0.1\,\rm \mu G)^2 \,\rm eV/cm^3$ by more than
3 orders of magnitude. For 
such a high total energy of relativistic electrons, with
the spectra extending beyond 100\,GeV energy (needed
for the interpretation of the radio fluxes), the fluxes of IC gamma-rays should 
extend to the region of high energy $\gamma$-rays. In the case of
power-law spectra in the radio region
with $\alpha_{\rm r}\simeq (1.1-1.2) $,  the extension of the IC fluxes 
with the same index from the X-ray to the $\gamma$-ray region predicts fluxes
observable for the future GLAST detector (see solid line in Fig.~2). For steep
spectra with $\alpha_{\rm r}\geq 1.3$, the predicted IC flux would be 
only marginally observable. However, because of the high gas density in 
the Coma cluster  
(up to $3\times 10^{-3}\,\rm cm^{-3}$ in its centre, Briel et al. 1992) 
the flux of bremsstrahlung $\gamma$-rays should be observed by GLAST in any case
(see Fig.~2). 

In conclusion to this section we note that the models invoking a single 
electron population as 
considered above may encounter a morphological problem. As pointed out by
Bowyer \& Bergh\"ofer (1998) the spatial extent of the EUV emission $\leq 
19^\prime$  
is significantly smaller
than the $\sim 30^\prime$ size of the 30\,MHz radio emission (\cite{Hen89}),
even though the target photon field for IC EUV emission is uniform.
This would mean that either the ICM magnetic field somehow increases
at larger distances from the core of Coma, which is very implausible,
or that radio emitting electrons are more spread out than less energetic
EUV emitting electrons. Although the latter possibility cannot be excluded
in principle for single electron population models, scenarios assuming 
a different origin of the EUV and radio electrons would allow different
spatial distributions of these electron populations more naturally.
In the following section we consider such a model.

\section{IC origin of the EUV flux in high magnetic fields} 

The conclusion that the magnetic field in Coma cluster is small, 
$B\sim 0.1 \,\rm \mu G$, is an unavoidable result of the assumption that
the observed excess X-ray flux has an inverse Compton origin. However, the
values of the magnetic field in Coma deduced from Faraday rotation 
and depolarization measurements are at a level of 
$B\geq 1\,\rm \mu G$ (\cite{Kim90,Fer95}), which is also 
typical for other clusters (\cite{Kro94}). 

If this is the case, then an IC origin of the excess hard X-rays 
in Coma is absolutely excluded. Indeed, the target photon field
for such IC radiation is well known. Therefore the total number 
of electrons $N_{\rm e}(\gamma)$ in the energy range 
$\gamma \sim 5 \times 10^3 - 10^4 $ is fixed. Because these energies are 
larger than $\gamma_{\rm r, 1}\simeq 3\times 10^3 B_{\rm \mu G}^{-0.5}$ 
for fields of the order of $1\,\rm \mu G$ (see section 2),
the synchrotron radiation of those `IC X-ray' electrons 
would be at frequencies $\nu > 10\,\rm MHz$, accessible for
radio observations. An increase of the magnetic field by one order of magnitude
compared with what is assumed in Fig.~1 would therefore result in 
overproduction
of the observed radio fluxes at least by two orders of magnitude.

Thus the IC interpretation of the hard X-ray excess should be abandoned and  
other explanations must be explored, such as bremsstrahlung emission 
of the {\it suprathermal } electrons 
(\cite{Kaa98,Ens98}), or perhaps invoking the thermal emission
of a multi-temperature plasma -- with temperatures extending beyond
10\,keV.  

However, for the excess EUV flux an IC origin can still not be excluded
if in the energy region from $\gamma_{\rm euv}\sim 300$ to  
$\gamma_{\rm r, 1} \sim 3000$ the energy distribution of the electrons would
decline extremely fast
 in order to reduce the total number of the electrons with 
$\gamma \geq \gamma_{\rm r, 1}$ by those  two orders of 
magnitude relative to a standard power-law
 $N(\gamma)$ with $\alpha_{\rm e}\geq 3$.
Basically, this implies a sharp cutoff in $N(\gamma)$ somewhere in the gap
between $\gamma_{\rm euv}$ and $\gamma_{\rm r, 1}$. This scenario 
then suggests that the population of relativistic electrons with
energies below a few 100 MeV,  responsible
for the EUV emission, should be different from the population of electrons
of higher energies producing the radio emission in Coma.

Here we suggest that the low energy electrons can be naturally connected
with the old population of relativistic electrons injected into 
the ICM more than a few Gyrs ago, and perhaps even as early as 
at the epochs of the initial starburst activity in the Coma galaxies.

 Powerful galactic winds due to early starburst activity in galaxy clusters,  
which appears to be necessitated
by the observed high iron abundance, could enrich the ICM also with
relativistic particles (\cite{Vol96}) and magnetic fields (\cite{Vol99}).
The possibility that low energy electrons responsible for the excess EUV 
emission in a galaxy cluster could be the relics of such a starburst epoch
has been considered by Sarazin \& Lieu (1998). 
Here we develop a detailed model for a 2-population origin of the relativistic
electrons in the Coma cluster. The model in particular suggests that the EUV
emitting electrons have a starburst origin, whereas the radio electron have been 
produced as secondaries during recent epochs.
A sharp radiative cutoff -- instead of a radiative break -- below 
$\gamma_{\rm r, 1}$ will be naturally 
produced in the spectrum of the relic starburst 
electrons, provided that the 
injection/acceleration of these electrons in the ICM    
has ended at some epoch $z > 0.05$ (corresponding to the
last $\Delta t > 10^9\,\rm yr$). 

\subsection{Model calculations}

Assuming for convenience of calculations an 
Einstein-de Sitter Universe ($\Omega_0=1$), the cosmological time is 
\begin{equation}
t(z)=\frac{2}{3 H_0 (1+z)^{3/2}}= 1.3\times 10^{10} (1+z)^{-3/2}
\,\rm yr\; ,
\end{equation}
chosing $H_0 = 50 \,\rm (km/s) / Mpc$ for the Hubble constant.   
Integrating then Eq.(1) over time, one easily finds that  
all electrons which have been injected into the ICM
at an epoch $z>0$ would at present be cooled down, due to IC losses, 
to energies below 
\begin{equation}
\gamma_{\rm cut} = \frac{310}{(1+z)^{5/2}-1}\;.
\end{equation}
If these electrons were not  reaccelerated at some 
$z^\prime < z$, their energy distribution should have a sharp
cutoff at $\gamma=\gamma_{\rm cut}(z)$. Note that for $z\sim 1$ or larger, 
$\gamma_{\rm cut}\propto (1+z)^{-5/2}$. This dependence readily follows
from the comparison of Eqs.(1) and (4), which indicates that 
most of the radiative cooling of the electrons, and thus the formation of
the cutoff in the energy spectrum, occurs at the early epoch as the
injection/acceleration of relativistic electrons in the ICM drops.   

The kinetic equation describing evolution of the energy distribution
of electrons $N\equiv N (\gamma,t)$ in time reads:
\begin{equation}
\frac{\partial N}{\partial t}\, = \, \frac{\partial}{\partial \gamma}(P N)
\, +\, Q\; ,
\end{equation}
where  $P\equiv P(\gamma,t)$ is the energy loss rate 
and  $Q\equiv Q(\gamma,t)$ is the source function of the electrons injected into
ICM. Assuming that the injection has started at some time $t_0$, 
the general solution to this equation can be written as 
(\cite{Pac73,Sar99}):
\begin{equation}
N(\gamma,t)=\int_{t_0}^{t} Q(\Gamma_{\gamma},\tau)
\frac{\partial \Gamma_{\gamma}}{\partial \gamma} \,{\rm d}\tau\; ,
\end{equation} 
where $\Gamma_{\gamma} \equiv \Gamma_{\gamma}(t,\tau)$ is the energy
of an electron $\gamma$ at time $\tau\leq t$, which can be found
from the equation
\begin{equation}
\frac{\partial \Gamma_{\gamma}}{\partial \tau}
= - P(\Gamma_{\gamma}, \tau)\; 
\end{equation}
with the initial condition $\Gamma_{\gamma}(t,\tau=t)=\gamma$.

The most important term in $P(\gamma,t)$ is 
connected with the radiative losses given by Eq.(1), first of all 
with the IC losses of the electrons in the cosmic MBR field, 
$w_{\rm mbr}\propto (1+z)^4
\propto [t/t(0)]^{-8/3}$ (using Eq.~4). Two other terms are connected with
the interactions of the electrons with the ICM gas. The
ionization (Coulomb) losses depend very weakly 
(logarithmically) on the  energy $\gamma$ of the relativistic electrons, 
and for our study they can be approximated with a good accuracy as\footnote{
Note that for a non-ionized medium the ionization loss rate is somewhat
smaller, by a factor of $\simeq 2.5$.} 
(\cite{Gin79}):
\begin{equation}
P_{\rm ion}\approx 4.1\times 10^{-5} \,n_{\rm gas}\,\rm yr^{-1}\; . 
\end{equation} 
The energy loss time for an electron with energy $\gamma$ then is:
\begin{equation}
t_{\rm ion}=\gamma/P_{\rm ion}\approx 2.4\times 10^{8}
\left(\frac{n_{\rm gas}}{10^{-4}\,\rm cm^{-3}} 
\right)^{-1} \,\gamma \;\rm yr\, .
\end{equation}
Thus electrons with $\gamma\geq 100$, which are responsible for 
the IC EUV emission, will not be affected by the Coulomb losses
if the ICM gas density $\overline{n}_{\rm gas}$
(the {\it average} during the time after 
injection) would not exceed $10^{-4}\,\rm cm^{-3}$. However at lower energies,
depending on $\overline{n}_{\rm gas}$,
the Coulomb losses may result in a significant flattening of $N(\gamma)$
compared with the initial 
injection spectrum. 

In calculations of the energy distribution of electrons 
$N(\gamma,t)$ we take into account also the bremsstrahlung
losses of the electrons. Note that these losses do not play any noticeable
role for the formation of $N(\gamma,t)$ because at all energies $\gamma$
they remain significantly smaller than the sum of the ionization and 
IC losses (see Sarazin 1999).  
 However bremsstrahlung becomes
an important process for production of the $\gamma$-ray fluxes (see 
Sections 2 and 4).

For the further calculations we approximate 
the injection spectrum of electrons $Q(\gamma,t)$ as a separable function of
variables, $Q(\gamma,t)\rightarrow Q(\gamma)\times q(t)$, with the 
energy distribution $Q(\gamma)$ given by Eq.(3), and 
with a time profile of the injection rate approximated in the form
\begin{eqnarray}
q(t)& = & ~~~~1~~~~~~~~~~,\; t(z_0) \leq t \leq t(z_1) \\
 & & [t(z_1)/t]^m~~~ , \; t(z_1)\leq t\leq t(0) \nonumber
\end{eqnarray}
Here the epochs $z_0$ and $z_1$ correspond to the beginning 
and to the end of an efficient injection of relativistic particles
(accelerated mostly at the early galactic wind termination shocks,
see V\"olk et al. 1996)  
into the ICM, so that the time interval 
$\Delta t_{\rm inj}=t(z_1)-t(z_0)$ has the meaning of a  
characteristic duration of the electron injection phase  
(`starburst/wind activity'). 

\subsection{Origin of the `EUV' electron population}

The electron energy distributions calculated for 
$q(t)$ declining with different indices $m$ 
are shown in Fig.~3. A stationary injection during 
$\Delta t_{\rm inj}=2\,\rm Gyr$ starting from
$t(z_0=1.5)\approx 3.3\,\rm Gyr$ is assumed. This results in the epoch
$z_1\approx 0.8$ when the injection rate has sharply declined. 
A significant cutoff or a `knee' in the electron distribution 
above $\gamma_{\rm cut}(0.8) = 93$ is formed
only if the injection
rate drops sharply, $m\ge 4$. Otherwise the electron injection   
currently remains high enough in order to fill up the knee with
recently accelerated electrons, resulting in a feature resembling
more an ordinary spectral break rather than a cutoff at
$\gamma\sim \gamma_{\rm cut}(z_1)$. Then, as it is seen
from Fig.~4 (solid curves, corresponding to $m=2$), in an ICM magnetic
field $B\geq 1\,\rm \mu G$ the synchrotron fluxes would significantly exceed
the radio fluxes observed, although the IC flux is still by an order
of magnitude less than the excess X-ray flux.

Such an extraordinarily fast decline of
$q(t)$ would actually mimic a  Gaussian type function, with a dispersion 
$\sigma_{\rm t} \simeq \Delta t_{\rm inj}/2$ significantly smaller than
the time interval $t(0)-t(z_{1})$ between the epoch $z\geq z_{1}$ 
of the maximum injection rate and the present time $t(0)$. Note that
this kind of
injection could be a quite appropriate approximation  
for the electrons injected 
into the incipient ICM at the epoch of early starburst activity.
 
\begin{figure}
\plotone{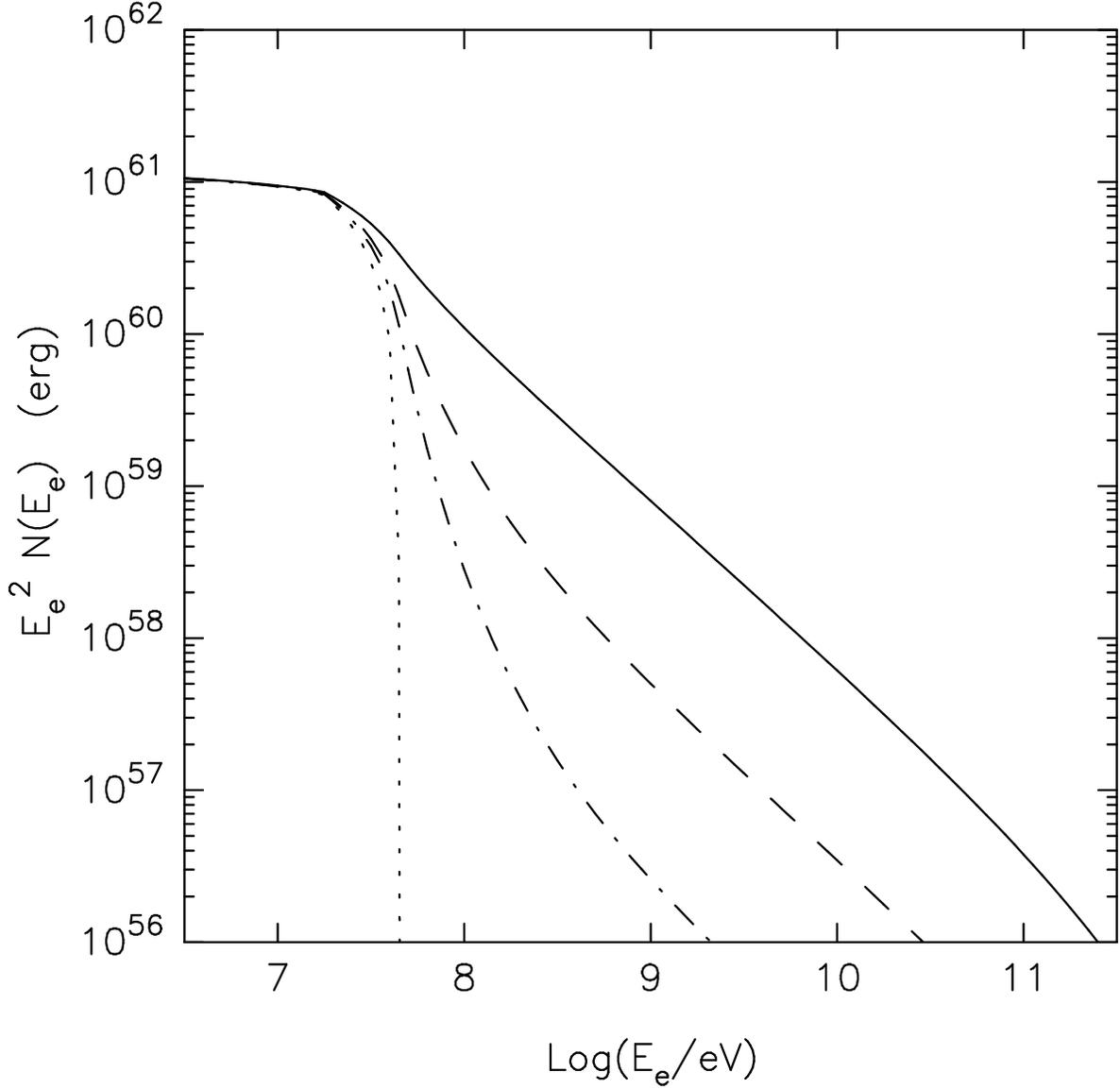}
\caption{The energy spectrum of electrons injected at the epoch $z\geq
z_0 =1.5$ during $\Delta t_{\rm inj} = 2\,\rm Gyr$ (i.e. $z_1=0.8$)
 calculated for an injection spectrum $Q(\gamma)$ 
with $\alpha_{\rm inj}=2.1$, and with a time profile of the injection 
rate given by Eq.(11), assuming different decline rates: $m =2$
(solid), $m=4$ (dashed), $m=6$ (dot-dashed), and $m=\infty$ (dotted curve).
In all cases the overall injected energy is $W_{\rm e}=10^{61}\,\rm erg$.    
It is assumed also that the gas density $n_{\rm gas}$ in the ICM at early epochs 
$z\sim 1$ was not much higher than the primordial baryonic density
at that time, $n_{\rm bar}(z=1)\simeq 2.4\times 10^{-6}\,\rm cm^{-3}$ (adopting
$n_{\rm bar}(0)\simeq 3\times 10^{-7}\,\rm cm^{-3}$). \label{fig3}}
\end{figure}

\placefigure{fig3}

\begin{figure}
\plotone{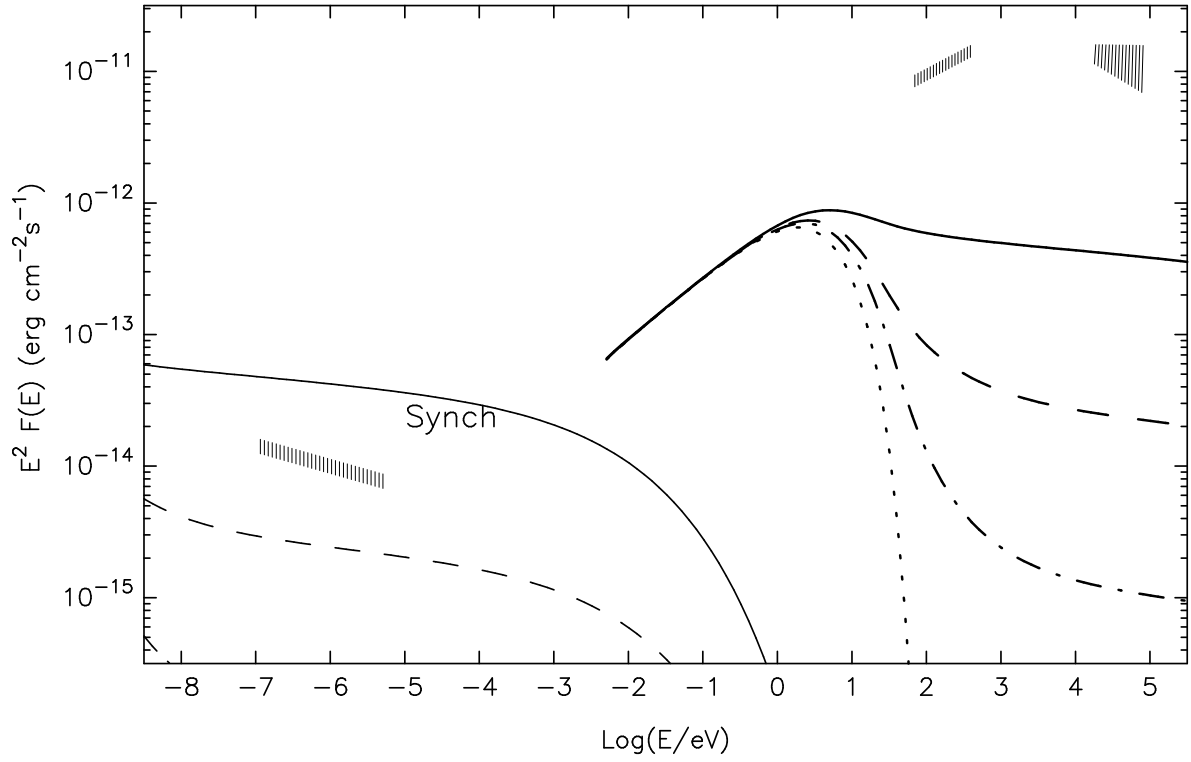}
\caption{The fluxes of synchrotron and IC radiations
produced by the electrons shown in Fig.3. A mean magnetic field
$B = 1\,\rm \mu G$ in the cluster is supposed. \label{fig4}}
\end{figure}

\placefigure{fig4}

The absolute values of the fluxes in Fig.~4 could be changed 
assuming a different
total energy for the injected electrons. However the cutoff in the IC radiation
would remain below the EUV region. Formally, the cutoff in the energy distribution 
of the electrons could be shifted to values $\gamma_{\rm cut} > 300$ needed 
for an explanation of the EUV flux, if we would assume that the phase of active 
injection of the electrons were continued until very recent epochs, 
$0.05\leq z_1\leq 0.2$,
but then it just stopped. However, this kind of assumption could hardly find
any physical justification, especially if this population of electrons is
due to the early starburst activity in Coma. 
 
Much more realistic seems to be a scenario where relativistic electrons 
driven into ICM by the galactic winds at the times of the early starburst activity
at $z_{\rm sb}\geq z_1\sim 1$, 
had cooled down to energies $\gamma_{\rm cut}(z_{1})< 100$,  remaining however 
relativistic. The number of these electrons, 
$N_{\rm e}^{\rm (rel)}\sim 10^{66}$ per each $10^{61}\,\rm erg$ of initial energy,
is very large. 
Then, the passage of a strong shock(s) across the cluster, 
e.g. due to `explosive' galaxy
mergers during the formation of the cluster (\cite{Cav92,Kon92}), 
would be sufficient to reaccelerate these 
electrons to GeV and higher energies. In the downstream region after the 
passage of the
shock the electrons would again be radiatively cooled down. A cutoff in the energy
spectrum of the electrons will then appear at energies  $\gamma \geq 500$ if
such a shock(s) occurred at $z\leq 0.2$ (i.e. during the last  
$\Delta t\leq 3\,\rm Gyr$).   

Another interesting possibility to `bring' the cutoff in the energy distribution
of the electrons into a `proper' place is connected with a purely adiabatic 
energization
of that relic population of electrons by compression of the ICM gas
due to contraction of the cluster. An interesting feature of the adiabatic
compression is that it does not change the shape of the initial 
energy distribution 
of the electrons, but just shifts their energy up by a factor of $a$: 
$N(\gamma) = N_{0}(\gamma/a)/a$. Therefore, after 
this kind of energization, the cutoff in the energy distribution
of the relic population would be conserved, and there is no need to assume
that the compression has stopped some $\Delta t \geq 10^9\,\rm yr$ ago
in order to allow the formation of the cutoff due to radiative losses
(the compression may go on even presently!).
For relativistic particles 
the parameter $a=(n_{\rm gas}/n_1)^{1/3}$, where  
$n_1$ and $n_{\rm gas}$ are the ICM gas densities before and after 
compression.
Assuming that the electron injection connected with the starburst 
activity in the cluster was basically over at say $z_1\simeq 1$, for 
the spectral cutoff formed at that epoch we would have $\gamma_{\rm cut}\sim
(60-70)$. Pushing this cutoff up by a factor of $a\simeq 10$, 
 we would explain the observed
excess EUV flux of Coma by IC radiation of the 
adiabatically reenergized relic 
population of electrons, but avoiding at the same time an  
overproduction of the radio fluxes at frequencies $\nu > 10\,\rm MHz$ 
(accessible for the radio observations of Coma) in a 
high ICM magnetic field $B\geq 1\,\rm \mu G$.

An adiabatic energy amplification factor $a\sim 10$ implies an increase 
of the gas density by a factor $\sim 10^3$. In principle, a compression
to even significantly larger factors is quite reasonable in the cooling 
flows of galaxy
clusters like A2199 (see \cite{Kaa99}).
 Although Coma appears to be a cluster without such a flow,
compression of the gas up to 3 orders of magnitude 
in its central region could be possible if the formation
of the gas density profile occurred not earlier than during the last
several Gyrs. Note that the angular
size of the excess EUV emission region (\cite{Bow98})
correlates well with the radius of the central core 
$\theta_{\rm c}\simeq 10^{\prime}$ of the (thermal) X-ray emission region
(\cite{Bri92}), where the maximum gas density is as high as 
$\approx 3\times 10^{-3}\,\rm cm^{-3}$,
and the mean gas density in the entire EUV emission region can be reasonably
estimated as $n_{\rm gas}\simeq 10^{-3}\,\rm cm^{-3}$. Thus, the parameter $a$
can be about 10 if the gas density profile in Coma 
started its formation (due to overall contraction, 
galaxy mergers) from $n_1\simeq 10^{-6}\,\rm cm^{-3}$. This would then imply
that the ICM gas in the core of the Coma has been increased to the level of
$10^{-3}\,\rm cm^{-3}$ not earlier than during the epochs $z\le 0.5$, because 
otherwise this initial value of $n_1$ were smaller than the primordial 
baryonic 
density. Indeed, given the primordial baryonic density at present
$n_{\rm bar}(0)\simeq 3\times 10^{-7}\,\rm cm^{-3}$, this density increases
as $(1+z)^3$ to $ 10^{-6}\,\rm cm^{-3}$ at $z=0.5$. Thus, the compression
timescales should then have been less than $t(0)-t(0.5)=6\,\rm Gyr$.   

\begin{figure}
\plotone{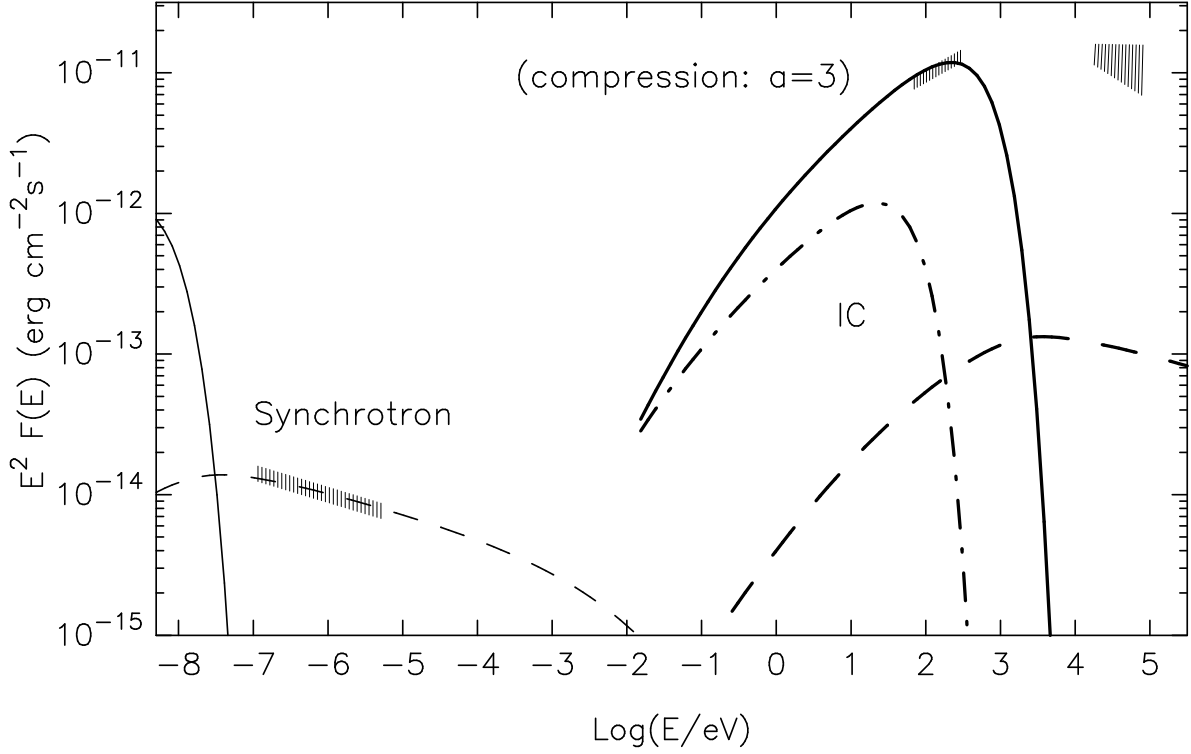}
\caption{The fluxes of synchrotron (thin lines) and IC (heavy lines) 
radiations
produced by the relic population of electrons injected into the 
ICM at epochs $z\ge 0.5$, but reaccelerated at $z_{\rm sh}\sim 0.5$, and 
later on compressed by a factor $a=(n/n_1)^{1/3}=3$ (solid lines). 
The spectrum of IC radiation calculated for 
the electrons without assumption of later  
adiabatic compression, is shown by the dot-dashed
line. The total energy of the electrons explaining the EUV flux is
$W_{\rm  e}^{\rm (inj)}=5.9\times 10^{61}\,\rm erg$. For the assumed
magnetic field $B = 1\,\rm \mu G$, the observed radio fluxes should be 
produced by a second population of electrons (dashed curve), e.g.  
assuming an injection with $L_{\rm inj}=2.9\times 10^{43} \,\rm erg/s$,
during the recent $\Delta t =1.7\,\rm Gyrs$ (corresponding to $z\leq 0.1$).
\label{fig5}}
\end{figure}

\placefigure{fig5}

It is important however that an essentially smaller  
degree of gas compression will be needed 
if we combine this adiabatic compression scenario with
the merger shock reacceleration at some $z^{\prime}< 1$.
In Fig.~5 we show an example of this kind of interpretation
of the excess EUV flux of Coma by IC emission (heavy solid line)
of the early population of electrons which have been (re)accelerated 
in ICM by the merger shocks at times $z\sim 0.5$, and later on compressed
by a factor $a=3$. This would require a compression of the gas only by
a factor of 27. Note that the gas density at radii 
$\theta \simeq 100^\prime$ (corresponding to $r\simeq 4\,\rm Mpc$)
inferred from the (thermal) X-ray fluxes is $\sim 10^{-5}\,\rm cm^{-3}$ 
(Briel et al. 1992), i.e. the gas compression in the central $\le 1\,\rm Mpc$
region of Coma up to a factor of $\simeq 100$ is quite plausible.

\subsection{Impact of ICM gas}

For calculations in Fig.~5, for the mean ICM gas density
{\it averaged in time} over the last $\Delta\approx 6\,\rm Gyr$ 
(corresponding to $z\leq 0.5$) we have assumed 
$\overline{n}_{\rm gas} =10^{-5}\,\rm cm^{-3}$. In that
case the Coulomb losses of the electrons become noticeable (assuming also
that the ICM gas was essentially ionized during that time) only at energies
$\gamma \le 10$, resulting in some flattening in the spectra of IC radiation 
below 0.1\,eV (see Fig.~5). Calculations show that the Coulomb losses 
have no a significant impact for interpretation of the EUV flux also for 
$\overline{n}_{\rm gas} \sim 10^{-4}\,\rm cm^{-3}$. However, the assumption
that the gas density in Coma was close to the current mean 
$n_{\rm gas}\simeq 10^{-3}\, \rm cm^{-3}$ during the recent epochs $z\leq 0.5$
dramatically changes the results. Indeed, as follows from Eq.(10), in that
case the electrons with $\gamma\leq 300$ would be effectively removed
by Coulomb losses. Thus, 
the IC origin of the excess EUV flux implies either that the increase of the
gas density in the central region of the Coma has occurred on timescales
no more than 2\,Gyrs (in the framework of the 
model invoking adiabatic energization), or that the relic electrons 
in the core have been reaccelerated by the shock(s) 
(1-2)\,Gyrs ago. Note that a shock propagating with a speed of a few 
1000 km/s travels 1\,Mpc during a time significantly less than 
1\,\rm Gyr.  In clusters with strong cooling flows, where the gas density
could reach values well in excess of that in Coma, it would be even
possible to expect {\it a deficit} of the excess EUV emissivity in the 
centre, if the infall timescale there would exceed the Coulomb loss time.

\subsection{Origin of the `radio' electrons}

The interpretation of the radio fluxes from the Coma cluster in the case of high 
magnetic fields requires a second component of relativistic electrons 
injected into the ICM at more recent times $\Delta t\leq 10^9\,\rm yr$. Because
the overall number of such electrons needed for production
of the observed radio fluxes is relatively small, the observed excess X-ray 
flux cannot be produced by those electrons (see heavy dashed curve in 
Fig.~5), 
and some other interpretation
(e.g. multi-temperature plasma emission?\,) has to be considered.  
At the same time, magnetic fields $B\geq 1 \,\rm \mu G$ allow  
a scenario where this second component of the electrons in Coma is due to
CR protons and nuclei interacting with the gas 
(via production and decay of $\pi^\pm$-mesons,  $pp\rightarrow 
\pi^\pm \rightarrow \mu^\pm \rightarrow e^\pm$), whereas 
such a secondary
origin of the radio emitting electrons in the case of low 
magnetic fields $B\sim 0.1\,\rm \mu G$ would require unrealistically
high energy in CRs (see \cite{Bla99}).
 The injection rate
of radio electrons assumed in Fig.~5 could be provided by relativistic
protons with a total energy 
$W_{\rm CR} \leq 3\times 10^{62}\,\rm erg$
(depending on the CR spectrum at energies $E_{\rm p}<100\,\rm GeV$). 
This seems to be a quite reasonable value, which can be  
compared with the energy $W_{\rm e}\simeq 10^{61}\,\rm erg$ assumed to be
injected into Coma in relativistic electrons initially (via galactic
winds), and with the energy $W_{\rm e}\sim 5 \times 10^{61}\,\rm erg$
in the relic electrons presently -- i.e. after their compression and/or 
reacceleration. 
Note that from the point of view of the secondary origin of radio 
emitting electrons in galaxy clusters it may also be meaningful that the Coma
cluster is characterized by a very high mean gas density up to 
Mpc spatial scales, comparable with the scales of the observed radio emission.  
The difference in the sizes of the EUV and radio 
halos can be perhaps connected with different spatial distributions of
high-energy relativistic protons and low-energy relativistic electrons. 
The reason is that the relic population of low energy electrons could
be more efficiently compressed (and adiabatically reenergized)
in the cluster core due to cluster contraction than the secondary
radio electrons produced currently from very high energy protons.
Both these protons and their secondary electrons can much more
effectively propagate across the cluster.

Thus, for high magnetic field scenario there is no need for current 
acceleration of particles in Coma at all. Moreover, an IC origin of the EUV
emission in that case allows current acceleration to occur only in 
some parts of the cluster, but not throughout of its volume,
because otherwise most of the `old' population of electrons would be involved
in acceleration, the cutoff above $\gamma_{\rm euv}$ would be removed, resulting
in a strong overproduction of the radio fluxes.

\section{Gamma-ray emission.}

An inverse Compton origin of the excess EUV emission requires an overall
energy in relativistic electrons with $\gamma\sim 300$ of order 
$5\times 10^{61}\,\rm erg$ in order to produce the observed flux. 
Given the high gas 
density in Coma, the expected fluxes of the bremsstrahlung $\gamma$-rays 
should be significant in both scenarios assuming low or high magnetic fields.
The solid curve in Fig.~6 shows the fluxes of bremsstrahlung $\gamma$-rays
produced by the relic population of the electrons which explain the EUV 
flux of Coma in Fig.~5.   

\begin{figure}
\plotone{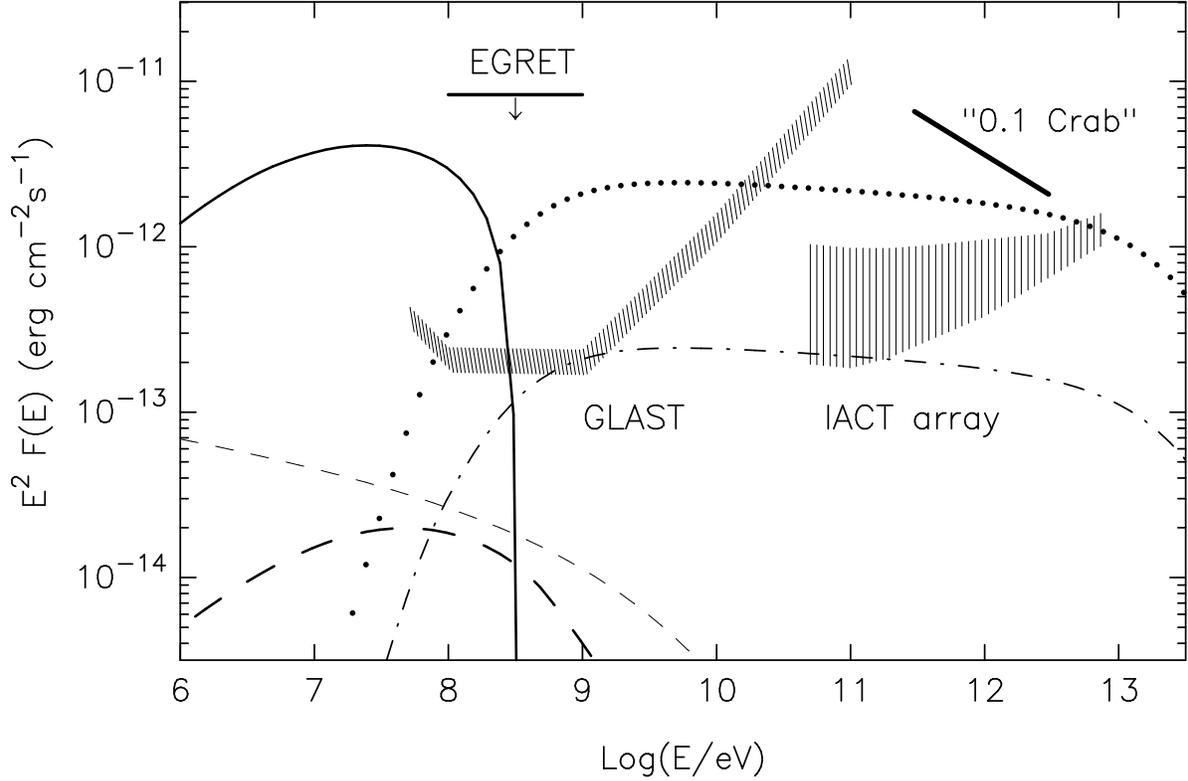}
\caption{$\gamma$-ray fluxes expected from the Coma cluster. The solid curve
shows the bremsstrahlung flux produced by the relic population of
electrons which explains the observed EUV flux of Coma in Fig.5.
Heavy dashed and thin dashed curves show the bremsstrahlung and IC fluxes
produced by the radio electrons (in the case of $B=1\,\rm \mu G$). Dotted
and dot-dashed curves correspond to the fluxes of $\pi^0$-decay $\gamma$-rays 
produced by CR protons with total energy $W_{\rm CR}=3\times 10^{62}$ and
$3\times 10^{61}$ erg, respectively (for $\alpha_{\rm pr}=2.1$).
The lower and upper edges of the shown range of sensitivities of IACT arrays
correspond to a point source and an extended ($\Delta \theta =1^\circ$) source,
respectively (from Aharonian et al. 1997). For comparison,
the flux level corresponding to
$10\%$ of the Crab Nebula flux (e.g. Konopelko et al. 1999)
is also shown. \label{fig6}}
\end{figure}

\placefigure{fig6}

A remarkable feature of the bremsstrahlung spectrum in Fig.~6 is 
the very sharp
cutoff at energies above a few 100 MeV. Comparison of this spectrum with
the $\gamma$-ray fluxes shown in Fig.~2 reveals that the scenario 
assuming
low magnetic fields in Coma, $B\sim 0.15\,\rm \mu G$,  can be easily 
distinguished from the case of high magnetic field by an essentially more
gradual decline (if any at all) of $\gamma$-ray fluxes above 300 MeV 
produced by GeV and higher energy electrons 
due to both the bremsstrahlung and the IC processes. An IC origin of
the EUV emission of Coma predicts in both cases $\gamma$-ray fluxes
well above the level of sensitivity of the future GLAST detector which thus 
would help to resolve the problem of low {\it vs.} high magnetic field
in Coma. This possibility to probe the strength of the magnetic field could 
be available also for other clusters with excess EUV emission
if their gas density would be high.    

Thus, the detection of a sharp cutoff in the spectrum of high energy $\gamma$-rays
by GLAST would confirm the $\geq 1\, \rm \mu G$ level of the magnetic field
in Coma. We note however that an absence of such a cutoff will not yet mean that
the magnetic field is low. At energies above 100 MeV one can 
expect also significant fluxes of $\gamma$-rays due to the production and 
the decay
of $\pi^0$-mesons if the energy content in relativistic protons in the central
$\leq 1\,\rm Mpc$ region of Coma is high, with  
$W_{\rm CR}\geq 10^{62}\,\rm erg$.
Fig.~6 shows that the total energy flux of the bremsstrahlung (solid curve) and 
$\pi^0$-decay (dotted curve) $\gamma$-rays may have even a single power-law
form if  $W_{\rm CR}\sim 3\times 10^{62}\,\rm erg$. 
On the other hand, Fig.~6 shows also
that in that case the fluxes of $\pi^0$-decay $\gamma$-rays should be
detectable also by the forthcoming imaging atmospheric Cherenkov telescope 
(IACT) arrays in the region of very high energies, $E\geq 100\,\rm GeV$
(hereafter VHE).
Note that the upper edge of the range of expected sensitivities of 
such IACT arrays  
(\cite{Aha97}) shown in Fig.~6 corresponds to an extended source
with angular size $\Delta \theta \simeq 1^\circ$, and the lower edge is
for a `point-like' source (with $\Delta t\leq 0.1^\circ$), and 
the expected angular size of the Coma cluster in VHE $\gamma$-rays
is $\leq 0.5^\circ$. Thus, in the case of a high CR proton content in Coma the
fluxes of the $\pi^0$-decay radiation, as well as the size of the emission
region, could be measured at both high and very
high energies. It is important to note also that 
at GeV and lower energies the spectral shape of $\pi^0$-decay radiation is not 
sensitive to the spectral distribution of CRs at very high energies.
Therefore in principle it should be possible to recover the spectral shape
of $\gamma$-ray fluxes of electronic origin in the $(0.1-1)\,\rm GeV$
region even if VHE emission of Coma were not detected.  

The measurements of VHE $\gamma$-ray fluxes can yield a rather accurate 
value for the mean cluster magnetic field, comparing these fluxes with the 
radio fluxes at high frequencies. This will be possible as far as the intensity 
of the secondary electrons will be then known, because they are
tightly connected with the $\pi^0$-decay $\gamma$-rays.   The angular
size of Coma $100\,\rm GeV$ should be comparable with the compact 
size of radio emission at GHz frequencies, with ${\rm FWHM} \sim 15^{\prime}$
(\cite{Kim90,Gio93,Dei97}).

\acknowledgements {We thank Felix Aharonian
for stimulating discussions. The work of AMA was supported through 
the Verbundforschung Astronomie/Astrophysik of the German BMBF 
under the grant No. 05-2HD66A(7)}


\clearpage

\clearpage

\clearpage

\clearpage

\end{document}